\documentclass[preprint,aps,amsmath,amssymb,superscriptaddress,nofootinbib]{revtex4-2}
\usepackage{hyperref} %
\hypersetup{ colorlinks=true, linkcolor=blue, citecolor=red,
filecolor=magenta, urlcolor=cyan }
\newcommand{\Case}[2]{{\textstyle \frac{#1}{#2}}}
\newcommand{\lP}{\ell_{\mathrm P}}
\begin{document}

\hfill\today

\title{Field Theoretic Aspects of Gravity\\ {\it From Direct Action to Quantum Fields}} 

\thanks{Contribution to Narlikar's Steady World: Man and the Legend}
\author{Ghanashyam Date} 
\email{ghanashyamdate@gmail.com}
\affiliation{Chennai Mathematical Institute,\\ H1, SIPCOT IT Park, Siruseri, Kelambakkam 603
103 India}
\begin{abstract}
	One of the technical anchors of JVN's research was the ``action-at-a-distance'' view
	of gravitational interaction, driven by ``Mach's Principle'' addressing the origin
	of inertia. This view exorcises the concept of a field as a mediator of force
	between separated ``particles''. However, within a quantum framework, the concept of
	field as a mediator transcends to field as an autonomous physical entity. Beginning
	with a comparison between direct action view and field theoretic view in classical
	framework and moving over to quantum framework, I sketch the strides made in quantum
	field theory including the new features brought in by gravity. This is intended as a
	conceptual trace rather than a review.
\end{abstract}

\maketitle\newpage

\tableofcontents 

\newpage
%
\section*{Introduction} 

The title is chosen to recall a series of small sized discussion meetings that were held
during 2000 till about 2016, supported by IUCAA, IMSc and SINP in the initial stages.  It
was Naresh Dadhich who proposed the title of the series. It was not until last year when I
and Sanjeev Dhurandhar wrote an introduction to the classic paper of Hoyle and Narlikar,
that I realized the `Field Theoretic Aspects' part of the title! For a high energy
physicist, born and brought up on (quantum) fields, the `field theoretic' was a redundancy.
Hoyle-Narlikar theory countered that ignorance. I thought it fitting to juxtapose the tussle
of ``fundamental field'' vs ``no field'' views as a tribute to both JVN and NKD. 

It is useful to have a sketch of a historical timeline in front to have a perspective on the
why of the struggle. After seeing how the two views emerge and attain their coexistence in
classical framework, we will switch to quantum framework in section IV. The subsections here
are organized around the principle conceptual shifts in the evolution of quantum field
theories and the novelties introduced by gravity. The aim is to trace main ideas in these
developments rather than to elaborate specific aspects. 

\section{A Historical timeline}

The Newtonian revolution introduced two distinct sets of laws: The Newton's laws of motion
and the law of universal gravitation.  The notions of inertia and force were introduced in
terms of which the laws of motion were stated. For distinguishing and linking kinematics and
dynamics, the postulation of ``inertial frames'' is essential.  The law of gravitation on
the other hand prescribes the gravitational pull a body exerts on another body. Both laws
have some gaps.  The laws of motion do not specify {\em how} an inertial frame is to be
identified and the question as to {\em why} is there inertia in the first place is left
open. The law of gravity gave a prescription but did not address how the gravitational pull
is actually {\em conveyed}.  To address the first question, Newton invoked {\em absolute
space} (1687) which was challenged by Mach (1883).  The second question was simply labeled
as ``action at a distance''. Not withstanding these gaps, the laws worked beautifully and
led to the world view that the world is fundamentally a {\em collection of particles} whose
dynamics is specified by the Newtonian forces. 

The idea of a ``field'' emerged when gravitational force was described in terms of a
gravitational field. The mathematical expression of gravitational field as the gradient of a
gravitational potential was introduced and mathematically developed by Lagrange (1773) and
Laplace during 1782-1799. Building on the empirical work of Coulomb (1785), this was adopted
for {\em electrostatics} by Poisson during about 1811-1813.  The works of Oersted, Ampere,
Biot and Savart during 1820-1830 extended the field representation to {\em magnetostatics}.
This was further elaborated by Faraday over the next couple of decades.  His lines of force
which can be mapped out experimentally made a convincing visualization.  This was a
convenient picture since the response of a body depended only on the ``local field''
regardless of the sources producing it. How a source produces the field was governed
separately.  

A major shift came when Faraday proposed that these fields are not just a mathematical
convenience but had a physical reality -- they modify properties of space.  Extending to
dynamical electromagnetic fields, Maxwell completed his equations (1861-1864) and predicted
electromagnetic waves which can carry energy and momentum. Thomson (1881) showed that
electromagnetic fields of moving charges carry energy-momentum and Poynting (1884) gave the
expression for energy-momentum flux of the fields.  The experimental demonstration by Hertz
(1988) followed.  Finally, Lorentz's work during 1892-1909 gave the force law providing
grounds for both particles and fields to coexist as realities. It was really the phenomenon
of electromagnetic radiation that strongly suggested the electromagnetic field is an
independent physical reality. This was supported by relativity theories of Poincare,
Einstein, Minkowski and quantum field theory of Dirac, Heisenberg-Pauli.

First doubts about the field concept were raised by Gauss (1845).  Imagining a ``medium''
pervading all space may be too much of a price to pay.  Why not have a direct action
formulation which nevertheless {\em propagates at finite speed}. Precisely this was achieved
by Schwarzschild in 1903 \cite{Sch1903}. His formulation had no instantaneous action of
Newton, followed special relativity and did not need any field. The next step in this
direction was taken by Tetrode in 1922 \cite{Tetrode1922}. The phenomenon of electromagnetic
radiation was already well established. From essentially the same action, he emphasized that
{\em an isolated particle cannot radiate} and asserted that ``If the Sun were alone in
space, it would not radiate.'' This challenged autonomy of field in radiation phenomena. In
1929, Fokker put the same action in a manifestly relativistic form and deduced the Lorentz
and the Maxwell equations (with sources) \cite{Fokker1929}.  In modern notation, the action
has the form,
\begin{eqnarray}
	S[x_a(\tau_a)] & := & -\sum_a m_a\int \sqrt{-dx_a^{\mu}dx_a^{\nu} \eta_{\mu\nu}} +
	\sum_{a < b} e_a e_b \int \int \delta( (x_a-x_b)^2)
	(dx_a^{\alpha}dx_b^{\beta}\eta_{\alpha\beta}) \ , \
\end{eqnarray}
where $a$ labels particle with mass $m_a$ and charge $e_a$.  $x_a^{\mu}(\tau_a)$ denote
their worldlines in a Minkowski background and the integrals are over worldlines. 

The variation is of the worldlines, $\delta x^{\mu}_a(\tau_a), \forall a$. Stationarity of
this action gives equations of motion for the particles and $A^{(b)}_{\mu}(x) := e_b\int
dx^{(b)}_{\mu} \delta( (x-x_b)^2)$ defining the vector potential due to particle $b$, at a
location $x$ satisfies the Maxwell equations with sources. This potential is half the sum of
the retarded and advanced Green's functions. This formulation of
Schwarzschild-Tetrode-Fokker (STF) is known as the {\em direct action} formulation.

Both the field theoretic and the direct action formulations worked fine for static and
steady-state electromagnetism. However, the phenomenon of electromagnetic radiation posed
challenges to both. In particular, the time symmetric fields arising in the STF formulation
will have no radiation or Poynting flux at infinity at all.

\section{The Radiation Reaction Problem} Accelerated charge emits radiation quantified by
Poynting flux far away. So the accelerated charge must loose energy-momentum and hence must
feel a reaction force.  The existence of such radiation damping was already noted by Hertz
in experiments with antennas. How does the damping force arise? 

{\em Field theoretic viewpoint:} A field theoretic mechanism begins by recognizing that the
field at the location of the charge is a sum of the fields due to all other sources {\em
and} the self field. The damping is almost instantaneous, so it can't be caused by far away
sources. And the only other available self-field is divergent for point charges. 

Nevertheless, during 1902 - 1905, Lorentz and Abraham, computed  a reaction force. They took
the charge to be a finite size sphere and computed the Lorentz force of one charge element
on another one.  Integrating over the charge sphere and expanding the retardation effects
they found that leading term in the force is proportional to the acceleration. So its
coefficient can be interpreted as {\em electromagnetic mass} and be included as a
contribution to the ``inertial mass''. This is divergent in the point limit.  The
sub-leading term, proportional to $\dot{\vec{a}}$, is finite in the point charge limit.
This is the radiation reaction force term. This leads to a third order equation of motion
for the charge which has runaway solutions and pre-acceleration, both are physical
pathologies.

Dirac \cite{Dirac1938} arrives at the same equation, for a point charge, but calculating the
energy flux through a thin world tube around the charge particle world line. On the world
line both the advanced and retarded solutions are divergent and away from it, they have a
$1/r^2 + 1/r$ fall off. The $1/r^2$ fall off is singular as the world tube is shrunk and
non-radiative.  The $1/r$ radiative piece is finite. So the natural proposal was to identify
half the retarded {\em minus} the advanced solution as the self field responsible for
generating the self force and interpret half the sum as contributing to the the
electromagnetic mass. Adding this to the (unknown) mechanical (or non-electromagnetic) mass
of the charge and interpreting the sum as the observed physical mass was the first instance
of ``renormalizing away divergences''. With this {\em classical} mass renormalization, the
point charge limit could be taken. 

Since the same equation is derived, the pathology of runaway solution and pre-acceleration
remains. In 1951, Landau-Lifshitz \cite{LandauLifshitz51} derived a reduced order equation
observing that the coefficient of the derivative of the acceleration in the LAD equation is
very small so may be treated iteratively. This avoids the pathologies.  When the coefficient
ceases to be small, the classical framework breaks down.

While the Dirac's choice is well motivated, it is mathematically driven without a clear
physical basis. Furthermore, the field theory is still silent on why only the retarded field
is seen in radiation. This is simply put in by hand. 

{\em Direct action viewpoint:} As noted above, the direct action of STF, produces time
symmetric fields which do not produce radiation far away from the worldlines.
Wheeler-Feynman \cite{WheelerFeynman} introduced the absorbers and sorted out both the
challenges posed by radiation.  There is no fundamental EM field in their theory.  The
theory was based on the four postulates: (a) An accelerated charge in empty space does not
radiate (Tetrode suggestions); (b) field acting on a charge arises only from the other
charges (no self field); (c) this field is one half the sum of the advanced and retarded
solutions (time symmetric) and (d) there are sufficiently many charges (absorbers) to absorb
completely the radiation emitted by any source. The first three postulates are already
contained in the STF action. The fourth postulate is what introduces {\em a fundamental
change}. They showed that with sufficiently many absorbers to ensure complete absorption, at
a point near an accelerated charge, the sum total of fields due to all absorbers equals half
the {\em difference} of the retarded and advanced fields generated by the accelerated
charge. The accelerated charge itself generates the time symmetric field which is half the
{\em sum} of the retarded and advanced field at the same point.  Hence, the total field is
just the retarded field due to the accelerated charge, thus explaining why only retarded
radiation field is seen. This feature is however dependent on the properties of the whole
universe -- it should be completely absorbing.  Furthermore,  the radiation damping force is
exactly the same as that given by Lorentz-Abraham theory.

Since, there is no fundamental self field, no self interaction, there is no infinite self
energy problem either.  As the same LAD equation is deduced, it has the same mathematical
pathology. However, the absorber assumption requires a self consistent boundary condition
which eliminates the runaway solution. The pre-acceleration survives, however this is just
the advance response from the rest of the universe and there is no physical pathology.  The
direct action formulation thus solves the key issues of radiation theory and (or but) also
brings in a role for the whole universe.

As far as classical electrodynamics is concerned, both field theory and direct action theory
maintain their rivalry.

\section{Machian formulation of relativistic gravity}
Since the Wheeler-Feynman theory showed that field degrees of freedom in electromagnetism
can be dispensed with, Hoyle and Narlikar sought to do the same for other fields.  In a
series of four papers \cite{HN-1964}, they developed a direct particle theory of
relativistic gravity and in the process also gave the most explicit implementation of Mach's
principle.  A brief summary may be seen in \cite{GD-SD-2025} and a very readable elaboration
in the cosmological context may be seen in \cite{JVN-1993}.  Here I just note some of the
differences from Wheeler-Feynman theory.

Focusing only on the gravitational interactions among masses, the action is written in terms
of world lines of particles in a form similar to the Fokker action,
\begin{equation}
	S_{new}[x(\tau_a)]  :=  \frac{\lambda}{2}\sum_{b\neq a}\int \int
	\tilde{G}(x_a,x_b)d\tau_a d\tau_b\ , ~ ~ d\tau_a^2 := -
	g_{\mu\nu}(x_a)dx_a^{\mu}dx_a^{\nu} ~ , ~ x_a := x(\tau_a)\ ,
\end{equation}
where, $\tilde{G}$ is the time symmetric Green's function for a scalar wave equation on the
curved space with metric $g_{\mu\nu}$ and is defined by,
\begin{equation}
	\big(g^{\mu\nu}\nabla_{\mu}\nabla_{\nu}\big)_x \tilde{G}(x,y) + \mu
	R[g]\tilde{G}(x,y) = - \frac{1}{\sqrt{-det(g)}}\delta^4(x,y)\ . 
\end{equation}
The choice of $\mu = 1/6$ makes the equation Weyl invariant.  

Define a {\em mass function} due to particle $a$ as,
\begin{equation}
	m^{(a)}(x) : = -\lambda \int d\tau_{a}\tilde{G}(x, x_a) ~ , ~ x_a := x(\tau_{a}) .
\end{equation}
Then $m_a(x_a) := \sum_{b\neq a}m^{(b)}(x_a) $ is the total mass of particle $a$ (along its
world line) contributed by all {\em other} particles. The action then has the same form as a
sum of the usual worldline actions, $S_{new} = -\sum_a \int d\tau_a m_a(x_a)$. The crucial
difference is that the mass parameter $m_a$ is a non-local function of all other particle
worldlines and is not a constant. {\em This is an explicit implementation of Machian
inertia}.  There is also an implicit dependence on the metric $g_{\mu\nu}$ through the
Green's function as well as through the proper time parameter.  Both the metric as well as
the world lines are {\em varied independently} in getting the equations of motion.

Varying the action with respect to the world lines, $\delta x^{\mu}$ keeping the metric
fixed, gives the equation for the world line which look similar to the usual geodesic
equation (but it is not) while varying the metric gives the field equations. These are the
equations (8) and (32) of the fourth paper in \cite{HN-1964}. 

The key points are that the world lines do not follow geodesics since the masses $m_a$ are
not constant. Secondly, the field equations have the structure that if there is no particle
or only a single particle, then the $m_a$ is zero and the field equations become $0 = 0$.
The whole action is identically zero. In this sense, even though the metric is varied
independently, it does not represent an independent degree of freedom (independent of
bodies). It is also true, that if one makes the {\em smooth fluid approximation} (which
includes proper response of the universe), the equations collapse to the usual Einstein
equations that have the usual field theory form in the Einstein frame \cite{JVN-1993}.

The HN theory, quite independent of its historical motivation and context, can be taken as a
specific non-local worldline based theory and analyzed for its solution space for a few body
universe. In the smooth fluid approximation it already resembles a field theory, a universe
of a compact binary should be more interesting. I am not aware of any attempts in this
direction.

{\em A Note:} When HN theory was proposed, the theory of gravitational radiation was just
beginning to get clarified. Now gravitational waves have been detected and the same twin
problems of the radiation reaction force and observation of only retarded fields also arise
in gravitational radiation. These are addressed in a field theory framework. The latter is
answered as an independent choice. The former has technical sophistication (compared to the
EM case a la Dirac) and is now understood as the Mino-Sasaki-Tanaka-Quinn-Wald (MiSaTaQuWa)
equation \cite{Poisson-2011}.  The gravitational case is much more complicated since there
is no analogue of local Poynting theorem.  Nevertheless, loss of energy-momentum in local
source dynamics, as obtained by including the reaction force does match with the balance
equations at future null infinity order-by-order \cite{MinoBarackPound}.

\section{Return to field theory (with the quantum)}
So far the perspective has been entirely classical and in retrospect, narrower. The problems
posed by the radiation phenomenon served to bring out this limitation. With the quantum
perspective, the whole understanding changes and along with it the perceived problems in the
classical framework. Now quantum fields take the center stage.
\subsection{Early understanding of radiation phenomenon}
Sticking to quantum electrodynamics for the context, the main changes have been two fold.
The classical trajectories of accelerated charges are replaced by only an averaged view.
Secondly, the classical field is taken to be real on par with particles to the extent that
it also gets quantized. At the fundamental level, the electromagnetic interaction stipulates
amplitudes for elementary processes of emission/absorption of photons by charged particles.
Under some conditions the amplitudes add up coherently while under some other conditions the
probabilities add up. When a charge is accelerated, the elementary bremsstrahlung amplitudes
add up producing a coherent state of the field. This manifests as the classical EM
radiation. Thus, classical radiation is not a fundamental autonomous entity but a state of a
quantum field. The momentum conservation implicit in the elementary processes also generates
a reaction force and impacts the averaged motion of the charge.  This gives rise to the LAD
force as an effective description. The effective description does not extend to the runaway
solutions or to the pre-acceleration implied in the classical description thereby
eliminating the older unphysical implications. For QED calculation of radiation reaction see
\cite{RadiationReaction-QFT}.

To realize the inverse process -- a charge absorbing radiation and getting accelerated -- we
can imagine putting in a charge inside a cavity containing radiation. The inverse
bremsstrahlung processes would be present, however depending on the coherence of the
radiation it may add the amplitudes or the probabilities. In the latter case, we do not get
a coherent motion of the charge. In the former case, the amplitudes would add and produce a
coherent motion. Why we do not see the time reverse of radiation is because such incoming
coherent radiation state is extremely rare. This rarity is seen locally in every region. And
an explanation of the rarity of such states is ultimately a property of the universe. 

The classical theory of Wheeler-Feynman sought the solution in a deterministic framework and
absorber mechanism was a physical way of implementing the solution. The QED has a stochastic
element in observation of phenomena and the explanation also pushes it back to a property of
the universe.

How does QFT handle the divergent self energy issue? Classically it was already sorted out
by the well motivated mass renormalization, including the specific procedure of Dirac. In
QED the issue resurfaces as UV divergences from the loops. Analogous to the logic of mass
renormalization, the parameters entering through the Lagrangian may be regarded as
auxiliary, to be modified by a renormalization process and finally identified with the
physically measured values. Thus QED has the necessary ingredients to answer the problem
posed by radiation phenomena.
\subsection{Extension to other field theories and renormalization}
We also have the weak and the strong interactions which are revealed only at the microscopic
level, without any classical manifestation. They are successfully described in a quantum
field theoretic framework. Do these more general QFTs bring out further new features? 

In a first step, consider extending QFT framework from QED to standard model. This still
uses the same {\em perturbative} computational tools as used in QED with one point of
departure. In QED, the unperturbed theory or the ``vacuum'' is naturally identified as the
Poincare invariant state. In standard model, there is choice of vacuum which spontaneously
breaks the $SU(2)\times U(1)$ symmetry. Thus not only is the action to be specified, a
vacuum state also needs to be chosen. After including the Higgs mechanism, the
renormalization procedure adapts well with no additional essential conceptual
feature\footnote{Technical issues such as dimensional regularization are not counted here.}.
The perturbative renormalization scheme already recognizes that the subtraction of
infinities has an arbitrariness in the choice of renormalization conditions.  Hence the
renormalized (or finite) masses and coupling parameters change as the renormalization scale
is changed. This leads to the generic features that the masses and couplings ``run''.  

However there are QFTs such as the four Fermi interaction theory of the weak interactions
and (perturbative) gravity for which this algorithm of perturbative renormalization fails.
These are the {\em perturbatively non-renormalizable QFTs}. Arbitrary scattering
amplitudes/cross-sections cannot be computed in terms of finitely many parameters such as a
few masses and couplings which can be inferred from experiments. At best, one could
interpret such theories as {\em effective field theories} which are valid below certain
cut-off momentum scale, $\Lambda$. If we are interested in processes with energy scale
${\cal E} \lesssim \Lambda$ and with a precision $\epsilon \gtrsim {\cal E}/\Lambda$, then
we need only finitely many terms in the action to be included. The renormalization process
then needs only finitely many parameters to be extracted from experiments and we have a
predictive model for calculation. Although the cut-off cannot be taken to infinity, we still
have the renormalization procedure and the feature of running parameters. 
\subsection{Defining a QFT a la Wilson} 
A conceptually significant advance was made by Wilson who shifted the focus from individual
processes to the underlying path integral or generating functional. In the process, he not
only gave alternative computational strategy, but also suggested a new view of what
constitutes a quantum field theory. His procedure involves constructing the (Wilsonian)
effective action by carrying out the path integration only over degrees of freedom,
$d\phi(k')$ with $k' > $ some $k$ and an integrand of the form $e^{-S_{eff}[\phi(k')]}$.
After a further integration from labels $k$ to $k-dk$, assuming locality, the new $S_{eff}$
has the same form as before but with different coefficients (masses and couplings). The
integration process relates the two sets of coefficients giving rise to a {\em
Renormalization Group} (RG) flow. Thus starting with some bare action, typically restricted
only by the field content and symmetries, as more and more degrees of freedom are integrated
over, the couplings change as per a coupled set of {\em renormalization group equations.}
The couplings thus encode quantum effects of the degrees of freedom which are below any
given the spatial resolution (or above momentum scale $k$). A QFT is now understood as being
determined by an RG flow on the space of all possible (local) couplings of a chosen field
content, a {\em theory space}. A mathematically more convenient presentation of this process
is given in terms of an {\em averaged effective action}, $\Gamma_{k}[\phi]$ by Wetterich
Exact Functional Renormalization Group equation (FRG) \cite{ExactRG}.  The effective action
is related to the Legendre transform of the Wilsonian effective action.  It is the
generating function of the usual vertex functions which denote the sum of 1-particle
irreducible diagrams of perturbation theory. There are technical differences between the two
procedures in how the integration is done, but the idea of QFT as being determined by an RG
flow in a theory space remains the same. The procedure is inherently non-perturbative and
there is no explicit mention of any vacuum state.

The RG flow is typically expressed in the form $k\Case{d g_i(k)}{dk} = \beta_i(g_j(k))$
where $k := \Lambda e^t, t \in \mathbb{R}$ denotes a variable scale and $g_i(k)$ are the
infinitely many, dimensionless coupling parameters. In taking $t \in \mathbb{R}$, we have
extrapolated the beta function equations to scales beyond $\Lambda$.  The simplest necessary
conditions for predictive QFT is to have fixed point(s), $\beta_i(g_{j*}) = 0\ \forall i$.
Linearization near the fixed point defines the matrix $\Case{\partial \beta_i}{\partial
g_j}\big|_{g_*}$ whose eigenvalues control how the flow moves near the fixed point as $t\to
\infty$.  An eigen-direction is UV attractive/repulsive if the corresponding eigenvalue is
{\em negative/positive}. Zero eigenvalues denote marginal directions and one needs to go
beyond linearization to study the flow. The UV-attractive directions are called {\em
relevant directions}. A theory is {\em asymptotically safe} if it has a {\em fixed point
with finitely many UV attractive directions} \cite{AsymSafety}. Such a theory is said to be
{\em UV-complete}. An {\em asymptotically free theory} is a special case of an
asymptotically safe theory with a trivial fixed point ($g_{i*} = 0\ \forall\ i$). How does
this help?

Consider an asymptotically safe theory. At finite values of $t$ or the scale $k$, the
initial couplings are near the fixed point. As the scale is {\em lowered}, the couplings
flow to low energy. We do not expect to reach another fixed point since at low energy we do
not have scale invariance. Nevertheless, the flow may become very slow as $t \to -\infty$.
These almost constant couplings can be determined by low energy experiments giving equations
for the initial values. And finitely many experiments suffice thanks to asymptotic safety.
Each RG flow trajectory, potentially singles out a specific model (= family of effective
actions) to be used in computations. It is worth noting that given a family
$\Gamma_{k}[\phi]$, the {\em quantum equation of motion}, $\Case{\delta\Gamma_k[\phi]}
{\delta\phi} = 0$, has several solutions depending on global and/or initial conditions. The
quantum vacua encoded as $\langle vac|\hat{\phi}|vac\rangle := \phi$, are contained in the
solution space of the quantum equation of motion. {\em Unlike perturbative QFT which
presupposes a vacuum, the FRG allows more general possibilities}.

While any specific flow is sensitive to the infrared regulator of Wetterich or integrating
out procedure of Wilson, the qualitative behavior is expected to be robust. Many
possibilities for the flows are conceivable. For physical interpretation and predictivity,
the asymptotic safety scenario suffices. With this, we have reached a general QFT
formulation and the conditions for selecting useful models. 

{\em Is gravity asymptotically safe?} Thanks to the works of Reuter and collaborators
\cite{AsymSafety}, the indications are that indeed Einsteinian gravity is asymptotically
safe with two or three relevant couplings (the dimensionless gravitational coupling and the
cosmological constant). The perturbative non-renormalizability of GR is not a fundamental
block as it was presumed to be.

\subsection{Gravity brings in diffeomorphisms}

So what new issues/features does GR bring in the QFT framework? It is the feature that the
background spacetime is dynamical which allows all possible diffeomorphisms of the manifold
on which the fields are defined and identifies the diffeomorphism equivalence classes of the
fields as relevant physical configuration space. To appreciate this let us briefly recall
how the computations are done.

The general computations in FRG framework for gauge theories commonly uses the so called
{\em background field method}. The (gauge) field is typically written as $A := A_{bgr} +
a_q$ where $A_{bgr}$ is a smooth background and the path integration is over $a_q$. All
computations, the action, the infrared regulator carry the dependence on the $A_{bgr}$.  The
computations are invariant under (infinitesimal) background gauge transformations so the
effective actions $\Gamma_k$ depend on the gauge equivalence class of $A_{brg}$.  There are
also {\em large gauge transformations} acting on the configuration space, those that are not
connected to the identity. The quotient of this full gauge group by the subgroup of the
infinitesimal gauge transformation is the {\em mapping class group}. By suitably summing
over the backgrounds related by large gauge transformation, we can expect to have the
effective action etc to depend only on the equivalence class of the background w.r.t. both
infinitesimal and large gauge transformation.  This decomposition of the configuration space
into gauge equivalence classes leads to superselection sectors of the quantum theory. The
$\theta$-parameter in the non-abelian gauge theories is an example of this.

The configuration space of the metrics on a given manifold also has similar decomposition
into {\em diffeomorphism} equivalence classes and the averaged effective action of the FRG
can be expected to be a function on the space of these equivalence classes (orbit space).
The mapping class group of diffeomorphisms is far more complicated and much less understood
than that of the non-abelian (Yang-Mills) gauge transformations.  There is no a priori
reason that these effective action functions on the orbit space be constant or suitably
`mild'.  Potentially then, the quantum gravity theories defined through asymptotic safety
framework, may be quite different on different equivalence classes.

To summarize, the FRG construction gives a viable quantum gravity (an RG trajectory)
dependent on the diffeomorphism equivalence class of metrics on a given manifold. These
quantum gravities (on the same theory space) could have qualitatively different properties
for distinct equivalence classes. This has a bearing on the issue of {\em background
independence} expected of a quantum gravity \cite{BkgrIndep}.  This is as far as the theory
space based on metric variables get us. We have reached another stage in QFT with GR
included.

Classical GR allows an equivalent alternative formulation in terms of tetrad variables
instead of a metric, in fact necessary to include spinor fields. An FRG type QFT can be
built with these fields.  However, a novel alternative exists. The tetrad based GR can also
be formulated in terms of a Yang-Mills connection and this paves a way for a {\em manifestly
background independent quantization}. Here is a brief digression.

\subsection{Background Independent Quantum Gravity}

Background independence here means that in the computational procedures no space-time or
spatial metric or any other non-dynamical field is singled out as a fixed background.  The
most developed scheme of background independent quantum gravity is canonical {\em Loop
Quantum Gravity} (LQG) \cite{LQG}.  Its basic phase space variables come from a 3+1
decomposition of the Hilbert-Palatini action which provides the densitized triad, $E^a_i :=
det(e)e^a_i$ while its canonical partner, the extrinsic curvature, combines with the
spin-connection to give an $SU(2)$ connection $A^i_a := \gamma K^i_a + \Gamma_a^i$.  For
these phase space variables, the parameter $\gamma$ has to be non-zero.  The Poisson bracket
between these two is given by, $\{A_a^i(x), E^b_j(y)\} = (8\pi
G\gamma)\delta^i_j\delta_a^b\delta^3(x,y)$.  The parameter $\gamma$ is the Barbero-Immirzi
parameter \cite{BarberoImmirzi}, conventionally positive while $G$ is the Newton's constant
which follows from the coefficient of the Hilbert-Palatini action.  The phase space has {\em
Gauss constraint} corresponding to the $SU(2)$ gauge invariance, {\em diffeomorphism
constraint} generating spatial diffeomorphisms and {\em Hamiltonian constraint} generating
the time evolution.  Holonomies of the connection along curves and fluxes of the densitized
triad through 2-surfaces, can be defined without any background 3-geometry and become the
natural choice of variables for a {\em background independent quantization}.  A canonical
quantization based on these variables leads to a well defined kinematical Hilbert space. One
of the crucial and qualitatively different implication is that the spectra of the
geometrical operators associated with areas, volumes, lengths are discrete in a specific
manner and depend on $\gamma$.  A complete quantization including imposition of the
constraints together with a satisfactory semi-classical phenomenology remains to be
achieved.  However, the two situations where a quantum gravity is expected to be essential,
LQG has demonstrated a promise. 

In the simplest context of the isotropic cosmological singularity a complete Dirac
quantization together with semi-classical regime has been carried out. In a specific way, as
the singularity is approached, the classical space-time begins to get obscured due to
quantum uncertainties, transits through a quantum regime and re-emerges as a classical
expanding universe. The discreteness of the quantum geometry is central to this resolution.
Secondly, in the context of stationary black holes modeled classically as quasi-local
isolated horizons, the quantum states are tied with the spectrum of the horizon area
operator.  Their counting gives the statistical mechanical entropy proportional to the
classical area together with computable quantum corrections \cite{LQGSuccess}.

The Barbero-Immirzi parameter so essential for the LQG quantization, can also be identified
as the inverse of the coefficient of the Nieh-Yan topological invariant\cite{NiehYan}.  It
drops out from classical equations of motion and has no classical manifestation.  In quantum
theory however, for every non-zero value of $\gamma$ the geometrical spectra are discrete
and it appears in the proportionality between the entropy and the horizon area. 

We have thus at least two distinct and very different constructions of quantum theories for
gravity, FRG and LQG both based on a field theory as distinct from quantum strings.  Of
course, a quantum framework is an extension of a classical framework and there can always be
multiple extensions. Internal consistency, containing the corresponding classical limit and
utility in extending phenomenology are the only broad requirements on any such extension.
Apart from the well acknowledged `quantum ambiguities', here we have examples of two
radically different quantum theories. How do we view them?

Let us note the features of these two which I believe are robust.

\subsection{Complementary Roles of FRG and LQG: A Curious Connection}

The FRG framework retains the continuum geometry feature of the classical GR based on
spacetime metric (or its Euclidean version). It has a scale dependent computational scheme
in terms of an RG trajectory in a theory space i.e. running couplings as an essential
feature. This is based on a {\em continuum geometry} (Euclidean to begin with!).  There are
strong indications that there is a non-trivial fixed point with 2 (and possibly 3) relevant
direction. The two low energy regime observables are the Newton's constant $G (\sim
G_{k=0})$ and the cosmological constant $\Lambda$ whose values can be used to select the RG
trajectory connecting to the fixed point values $g_*, \lambda_*$. The numerical values at
the fixed point vary somewhat depending on the regulator, truncation used, matter content
etc. The indicative values are: $(g_*, \lambda_*) \sim (0.85, 0.25)$ \cite{Values}. While it
can address phenomenology at arbitrary energy scales $E$ using $\Gamma_{E}$, it has no
obvious provision to model any micro-states of black holes. It also has potentially
non-trivial dependence on the diffeomorphism equivalence classes in the space of metrics.

The background independent LQG is based on different classical fields and has {\em discrete
geometry} as its robust feature. Its gravitational sector has {\em two} undetermined
parameters $(G_{LQG}, \gamma)$.  While it can model micro-states of black holes and provides
a bounce mechanism to resolve the isotropic singularity, its connection to low energy
phenomenology is tenuous. Being a background independent quantization, there is no natural,
variable scale that can lead to running of coupling parameters. It has thus no provision for
{\em direct access to determine $G_{LQG}$} and hence {\em $G_{LQG}$ need not equal $G$} as
suggested by the classical derivation. It is an undetermined parameter.  

Is there a common situation where both theories can make a prediction?  Interestingly, black
hole entropy provides a common context. 

Recall that black hole thermodynamics, deduced within classical theory led to the black hole
entropy as $S = A/(4G)$ after Hawking deduced the temperature to be $\kappa/(2\pi)$ invoking
quantum effects. All steps here use continuum geometry. The {\em statistical interpretation
of entropy} requires identification of states and their counting. In LQG, the states are
identified as the simultaneous eigenstates of operator form of the quasi-local horizon
boundary condition. This involves the area operator corresponding to the horizon. The
counting leads to the entropy as $S_{LQG} = [\Case{\gamma_0}{\gamma}] [\Case{G}{G_{LQG}}]
(\Case{A}{4G}) + o(A^0) + \cdots $ where $\gamma_0 (\approx 0.23753..)$ is the numerical
constant obtained from the asymptotic estimate of the state counting for large back holes
\cite{LQGSuccess}.  Since for large black holes, continuum picture is well justified, the
LQG entropy must match the Bekenstein-Hawking entropy. This gives us one condition on the
two undetermined LQG parameters. 

As noted above, FRG has no natural way of counting micro-states. Hence it cannot provide a
{\em statistical mechanical interpretation} of BH entropy. FRG with asymptotic safety
naturally provides an asymptotic value of running gravitational coupling, $G_{asy} := g_*G$
where $g_*$ is the dimensionless UV fixed point value.  UV regime is where quantum gravity
is anchored in FRG. The $G_{asy}$ is then a natural choice for $G_{LQG}$. While $G_{LQG}$
itself does not run, along an RG trajectory emanating from near the UV fixed point with
$G(k) \approx G_{asy}$, does run down to the Newton's constant $G$. 

Thus we may propose that scale dependent phenomenology of quantum gravity may be accessed
using FRG while LQG be taken as anchoring the UV complete form of quantum gravity. For this
to be consistent, the LQG parameters be identified as $G_{LQG} = g_*G, \gamma = \gamma_0/g_*
$.  This determination of $\gamma_{LQG}$ is well known, apart from the $g_*$ and
numerically, this identifications make virtually no difference.  Conceptually, however the
rationale goes beyond numerical identification. This suggests that FRG and LQG can be
distinct quantum theories of gravities best suited for scale dependent phenomenology and for
extension of classically indicated singular regimes respectively.

As an aside: Could we not expect $\gamma$ to be similarly identified with a fixed point
$\gamma_*$ of an RG flow?

Interestingly the search for a possible running $\gamma$ has been explored within the FRG
framework with Hilbert-Palatini action together with Holst/Nieh-Yan terms included
\cite{RunningBI}. While $\gamma^{-1}$ as the coefficient of the Nieh-Yan term ensures its
invisibility at the classical level, the relative coefficient of the two terms of the
Nieh-Yan invariant is not protected by any symmetry and the two terms flow independently
under renormalization. So far no robust hint of a fixed point behavior of $\gamma(k)$ has
been seen. The absence of a running $\gamma$ is perhaps a hint that {\em $\gamma$ is a new
fundamental constant necessitated by discrete quantum geometry.} Thankfully its
determination via the BH entropy also suggests transitioning to continuum geometry. 

The distinguishing feature of the two frameworks is the manifest background independence of
the quantum framework.  This is analogous to the feature of indistinguishability dividing
the frameworks of quantum and classical statistics.  Just as quantum statistics has to be
invoked when indistinguishability arises, so is background independence needed when the
fluctuations in the geometry induced by matter quantum fluctuations become large (so that
``background looses its meaning'').  We do not have to require that the LQG framework should
go over to the continuum geometry framework with its RG flow just as we do not need to lift
the (anti)symmetrization constraint on quantum states in a continuous manner to go over to
the Maxwell-Boltzmann statistics.  The need for indistinguishability is indicated by the
{\em degeneracy parameter}, $n\lambda_{th}^3$ for example.  Likewise, the geometrical area
in units of the flux quantum $\gamma\lP^2$ would signal the need for background
independence.

This concludes my conceptual trace of {\em quantum} field theory. Classical conception of a
field that appeared disposable becomes vastly richer in the quantum conception.

\section{Summary}

Evolution of the concept of a field has been our main focus. Let us summarize now. 

We saw how the acceptance of a field as an autonomous entity was neither obvious nor
unchallenged. The electromagnetic radiation phenomenon seriously challenged field theory and
brought in focus the direct action theories. This could even be extended to a gravity theory
incorporating Mach's principle most directly. However, the huge success of QED brought the
focus back to field theory, albeit in its quantum form. 

The QFT framework itself underwent significant conceptual changes from perturbative
renormalizability as a selection criterion, to more general effective QFTs, and finally to
the Wilsonian understanding of a QFT as a suitable RG flow in a theory space.  Weinberg's
Asymptotic safety criterion gives our most general characterization of renormalizable QFTs. 

Adding gravity brings in the additional feature of diffeomorphism invariance vastly more
complicated that non-abelian gauge invariance.  Potentially suggesting qualitatively similar
but distinct RG trajectories for different equivalence classes of spacetimes.
Interestingly, GR also affords a manifestly background independent quantization scheme which
reveals the novel possibility of discrete geometry. This plays a crucial role in a bounce
replacing the isotropic big bang singularity as well as providing direct state counting for
BH entropy.

These multiple ``quantum theories of gravity'' are {\em complementary} in their scope. One
does not have to be subsumed in the other. One can transit from one to the other when needed
as the example of quantum statistics and Maxwell-Boltzmann statistics shows. This suggests
that manifest background independence is akin to indistinguishability of backgrounds.
\begin{acknowledgments}
I acknowledge the use of ChatGPT-5.5 (OpenAi)  and Google Gemini-3.5 as sounding boards
during the preparation of the article.  This work is partially supported by Infosys
Fellowship.
\end{acknowledgments}
\newpage
\end{document}